\documentclass[a4paper]{aa}
\usepackage[varg]{txfonts}
\usepackage[english]{babel}
\usepackage[utf8]{inputenc}
\usepackage{graphicx}           
\usepackage{color}   %
\usepackage{natbib}
\usepackage{threeparttable}
\usepackage{booktabs, dcolumn}

\bibpunct{(}{)}{;}{a}{}{,} 

\title{Why is solar cycle 24 an inefficient producer of high-energy particle events?}

\author{Rami Vainio\inst{1} \and Osku Raukunen\inst{1} \and Allan J.\ Tylka\inst{2} \and William F.\ Dietrich\inst{3} \and Alexandr Afanasiev\inst{1}}
\institute{Department of Physics and Astronomy, University of Turku, FI-20014 Turku, Finland \and Emeritus, NASA Goddard Spaceflight Center, Greenbelt, MD 20771, USA \and Consultant, Prospect Heights, IL 60070, USA}
\date{Received date / Accepted date}

\authorrunning{Vainio et al.}
\abstract {} {The aim of the study is to investigate the reason for the low productivity of high-energy SEPs in the present solar cycle.} {We employ scaling laws derived from diffusive shock acceleration theory and simulation studies including proton-generated upstream Alfv\'en waves to find out how the changes observed in the long-term average properties of the erupting and ambient coronal and/or solar wind plasma would affect the ability of shocks to accelerate particles to the highest energies.} {Provided that self-generated turbulence dominates particle transport around coronal shocks, it is found that the most crucial factors controlling the diffusive shock acceleration process are the number density of seed particles and the plasma density of the ambient medium. Assuming that suprathermal populations provide a fraction of the particles injected to shock acceleration in the corona, we show that the lack of most energetic particle events as well as the lack of low charge-to-mass ratio ion species in the present cycle can be understood as a result of the reduction of average coronal plasma and suprathermal densities in the present cycle over the previous one.} {}
\keywords{Acceleration of particles -- Shock waves -- Sun: activity -- Sun: particle emission}

\begin{document}

\maketitle

\section{Introduction}
Solar energetic particle (SEP) events are outbursts of high-energy particles, mainly protons and electrons, from the Sun. Solar energetic particle events in the present solar cycle (SC) 24 and the previous cycle (SC 23) have been observed with high-fidelity instruments from space and with traditional ground-based instruments, i.e., the neutron monitor (NM) network. Both to the benefit and disadvantage of the solar research community, SC 24 has been unusually quiet in many respects. On the one hand,  the two most recent cycles have very different solar activity levels, which makes it  possible to identify their key differences and discuss the possible causal relations between the various manifestations of solar activity. On the other hand,  the present cycle is very quiet, so we do not have many extreme events to compare with the previous cycle. For example,  the full SC 23 hosted sixteen SEP events producing a ground-level enhancement (GLE), while the number of GLEs in the current cycle is only one.

The purpose of this paper is to briefly review the reported differences in SCs 23 and 24 and to use the  modeling results of SEP acceleration at the Sun to understand the reason for these differences. Instead of detailed modeling of individual events, we concentrate on the SEP climatology, i.e., on the differences in the typical coronal particle acceleration conditions as a possible explanation for the differences in the SEP cycle, including the properties of the ambient plasma and the coronal mass ejections (CMEs) driving fast shocks through the coronal plasma. Our modeling approach is thus to assume that the high-energy particles are accelerated by coronal shocks, and we base our assessment on scaling laws of SEP spectral parameters established by the theory of diffusive shock acceleration (DSA) \citep{bell78} and some recent simulation results \citep{vainio14,afanasiev15}.

Many of the previous studies \citep[e.g.,][]{gopalswamy14,gopalswamy14b,gopalswamy15,mewaldt15,makela15} comparing the particle production of the two cycles have highlighted the importance of the coronal magnetic field intensity as the controlling factor of the shock acceleration rate. We  challenge this interpretation, claiming that at least in a quasi-parallel shock geometry the maximum energy achieved by DSA is not controlled by the magnetic field intensity because of the self-generated Alfv\'en waves in the ambient coronal plasma. Rather, the DSA process seems to point towards the importance of the plasma density and especially the density of the seed particles (thermal and/or suprathermal) in the ambient medium.

\section{Review of key observations}
\subsection{Solar activity: sunspots}
\begin{figure}
\centering
\includegraphics [width=0.99\columnwidth, angle=0]{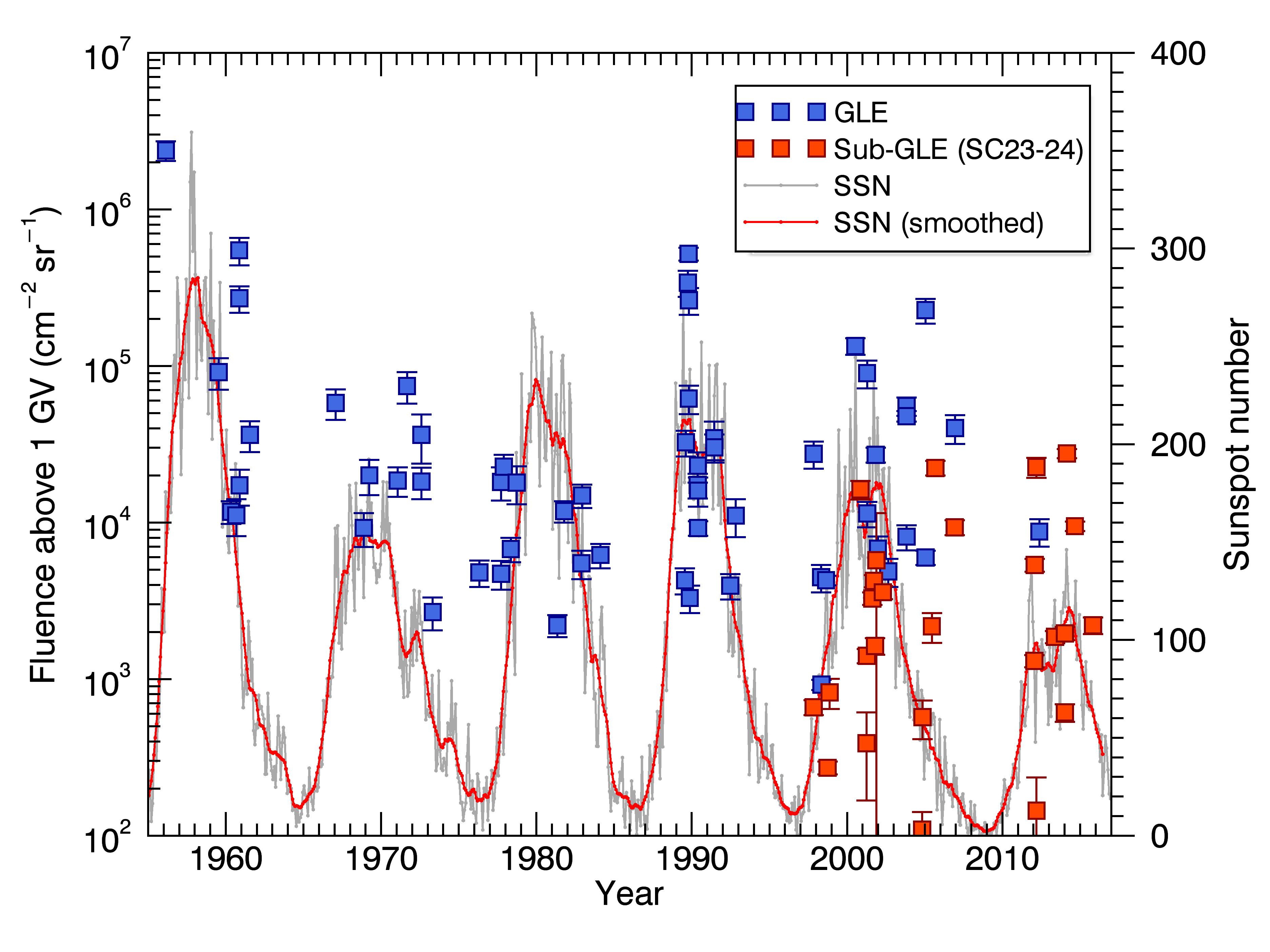}
\caption{ Monthly mean (gray) and the 13-month gliding averaged (red) sunspot number and the $>$1-GV fluences of all GLEs observed since 1956 (blue boxes) for which the NM statistics allows the determination of the spectral parameters of fluence. The $>$1-GV fluences of sub-GLEs of SC 23 and 24 (see \S \ref{sec:large_sep}) are included (red boxes) as well.}\label{fig:glecycles}
\end{figure}

The evolution of the solar activity since the 1950s, as measured by the sunspot number, is given in Fig.~\ref{fig:glecycles}.\footnote{Based on  SILSO data, Royal Observatory of Belgium, Brussels, available at http://www.sidc.be/silso/} While there is no consensus for the  definition of the solar-minimum conditions in terms of sunspot numbers, we can see that the present cycle is already well into its declining period. The plot also shows  the overall weakness of the present cycle in terms of sunspots, both when considering the peak value of the sunspot number and its time-integral over the solar cycle.

\subsection{SEP events and the related CMEs}
Several statistical studies have been performed comparing the SEP events of  SCs 23 and 24. The most notable difference in the events is the almost total lack of the most energetic events, the GLEs \citep{gopalswamy14,gopalswamy15}. Figure~\ref{fig:glecycles} shows the event integrated $>$1 GV proton fluence of all GLE events since 1956 (SCs 19--24), for which the spectral parameters and their error estimates can be determined using the method of \citet{tylka09}. The fluence values and their uncertainties are obtained by a Monte Carlo analysis (see \S\ref{sec:large_sep} below). When comparing the number of GLEs we consider the same number of days since the previous solar minimum defined as the 15th of the month when the 13-month averaged sunspot number attains the minimum. The date corresponding to the end of the considered time period in SC 24 (15 December 2016) would thus correspond to 15 May 2004 in SC 23. The total number of GLEs from 15 May 1996 to 15 May 2004 was 13, while the total number of GLEs in the present cycle, as noted above, is 1.\footnote{There was a second SEP-event related enhancement observed by two NMs on 6 January 2014 \citep{thakur14,kuehl15}, but as the monitors are both located at the South Pole, this particle event does not meet the official definition of a GLE.}

Not only is the number of GLEs in SC 24 significantly lower than during the previous cycles, but  the total fluence generated by these events also falls almost two orders of magnitude short of the fluence during either of the two previous cycles. The total fluence of protons in SEP events at lower energies was analyzed by \citet{mewaldt15}. They found that the accumulated fluence of $>$10-MeV ($>$100-MeV) protons during the first 2100 days of SC 23 was  a factor of 4.4 (6.4) higher than during the first 2100 days of SC 24. While the difference is not as dramatic as in the case of fluence related to GLEs, the trend is similar: the higher the particle energy, the higher  the contrast between the fluence during the last two cycles. However, by far the most dramatic effects in the Sun's particle accelerators are observed at the most energetic end of the spectrum.

\citet{gopalswamy14,gopalswamy14b, gopalswamy15} studied the properties of CMEs related to SEP events in the two cycles and found that while the total CME rate was very similar in the ascending and maximum phases of the two cycles, CMEs related to the SEP events in the present cycle were faster and more expansive. They noted that this implies that the coronal conditions in the two cycles are markedly different. \citet{makela15} found that the average distance of CMEs related to major non-GLE SEP events at the time of metric type II burst (i.e., shock) onset was somewhat lower in SC 24 than in SC 23, which supports the ability of the present-cycle CMEs to drive shocks through particle-accelerating coronal environments.

The heavy-ion abundances compared between the two cycles also support the picture of the lack of high-rigidity particles. \citet{mewaldt15} considered the ratios of abundances of the two cycles and found a clear scaling law showing that elements with the lowest charge-to-mass ratio, $Q/M$, are most depleted in SC 24. The SEP ion abundances relative to hydrogen  observed in SC 24 over that observed in SC 23 was proportional to $(Q/M)^{0.66}$. \citet{raukunen16} considered the daily intensities of the time periods with statistically significant amounts of heavy ions during the two cycles and found that the fluence distributions show the largest lack of high fluences for the low $Q/M$ elements. Thus, elements with the lowest $Q/M$  consistently show the largest depletion from SC 23 to 24.

\subsection{Ambient coronal and solar-wind properties}
In addition to the properties of SEP events, \citet{mewaldt15} studied the differences of the SEP-related phenomena during the first 5.8 years of the two cycles. They reported that the magnetic field, the plasma density, and especially the density of suprathermal seed particles was lower in SC 24 than in SC 23. \citet{gopalswamy14} analyzed the first 62 months of the two cycles and found similar scalings for the plasma parameters as \citet{mewaldt15}: total pressure, magnetic field, density, and ion temperature were all lower during SC 24 than during SC 23. We note, however, that the Alfv\'en speed had not dramatically changed from one cycle to the next \citep{gopalswamy14}.

\section{Review of modeling results}

\citet{vainio14} performed a parametric study of coronal and interplanetary shock acceleration of protons using the Coronal Shock Acceleration (CSA) code \citep{vainio07,vainio08}, which is a Monte Carlo simulation treating the upstream Alfv\'enic turbulence and ions self-consistently in terms of their energy exchange. The study established a scaling law between the cutoff proton momentum achieved through DSA and the parameters of the shock, the ambient plasma, and the seed particle density in the upstream medium,
\begin{equation}
p_c^{\beta} \approx p_{\rm inj}^{\beta}\frac{\pi(\beta+3)n_{\rm seed}}{4n}\frac{\Omega_{\rm p} r}{v_{\rm A}}\label{eq:vainio14}
,\end{equation}
where $\beta = 3r_c/(r_c-1)-3=3/(r_c-1)$, $r_c$ is the scattering center compression ratio of the shock, $p_{\rm inj}$ is the injection momentum, $n_{\rm seed}$ is the seed-particle density (i.e., the number density of particles at momenta higher than $p_{\rm inj}$), $n$ is the plasma density, $\Omega_{\rm p}$ is the proton cyclotron frequency, $r$ is the radial distance from the Sun, and $v_{\rm A}$ is the Alfv\'en speed of the ambient medium. This scaling law originates from the DSA theory of \citet{bell78} for the foreshock wave intensities. We note that contrary to an intuitive idea that the cyclotron frequency (and therefore the magnetic field) would scale the cutoff momentum, the dependence on it cancels out because of the Alfv\'en speed in the denominator of the equation. This is a consequence of the self-consistent amplification of Alfv\'enic turbulence in the upstream region. We also note  that the dependence on shock velocity is hidden in the use of radial distance as the variable. (The faster the shock, the higher the distance  achieved by a shock in a given time.) Since we are interested in the total number of particles accelerated by the shock, distance is a more suitable variable than time.

Limited acceleration time is not the only thing that can produce a break in the accelerated particle spectrum. Another factor limiting the acceleration to the highest energies in a fully time dependent model is the finite growth time of the waves resonant with particles at the highest energies exceeding the available acceleration time \citep[e.g.,][]{ng08}. Alternative mechanisms suggested for the breaks include the focusing-driven escape of particles to the upstream at high energies \citep{vainio00,vainio14}, the adiabatic cooling rate becoming comparable to the energy gain rate at the shock \citep{vainio00}, and time-dependent shock geometry with the quasi-perpendicular intitial phase producing the high-energy population with lower intensity than the later quasi-parallel phase not reaching the highest energies \citep{tylka06}.  Another way to produce broken power-law fluence spectra is through considering transport effects modifying the source energy spectrum, either within a test-particle approach employing time-integrated Parker equation \citep{li15} or within the self-generated wave model \citep{vainio03}. The present modeling approach is therefore not a unique choice for a basis of modeling spectra in large gradual events.

For strong shocks, $\beta\approx 1$, the scaling law for the  cutoff momentum of Eq.~(\ref{eq:vainio14}) can be written as
\begin{equation}
p_c \propto p_{\rm inj}\frac {r^2 n_{\rm seed}}{r_0^2 n_0}\sqrt{\frac{r_0^2 n_0}{r^2 n}},
\end{equation}
where $r_0$ is a reference distance and $n_0$ is the plasma particle density at that distance. Thus, assuming that shocks accelerating large SEP events are strong and have similar Mach numbers from one cycle to the next, we conclude that the differences in the cutoff momenta of the events are related mainly to the differences in seed-particle densities and ambient plasma densities.

The cutoff momentum is determined by the resonant interaction between the particles and plasma waves. Thus, it is actually  setting the limit on rigidity ($R=p/q$, i.e., momentum per charge) rather than momentum. Using shock acceleration theory, \citet{zank07} predicted that the cutoff in the proton momentum spectrum would be related to a sharp low-wavenumber cutoff of the Alfv\'en waves, which in turn would mean that heavy ions would have a cutoff in their spectrum at constant rigidity. \citet{battarbee11} considered the acceleration of heavy ions at shocks using the CSA code and noted that the scaling of cutoff energies per nucleon of different species followed a power law close to $E_c/M\propto (Q/M)^{1.5}$ to $(Q/M)^{1.6}$, which would correspond to cutoff rigidities scaling as $R_c\propto (Q/M)^{-0.25}$ to $(Q/M)^{-0.2}$. However, their study also showed that when the proton acceleration is more efficient,  the scaling is closer to the constant cutoff rigidity scaling law suggested by \citet{zank07}.

The CSA model and the DSA theory of \citet{bell78}, however, have one drawback that needs to be assessed: they employ a simplified resonance condition between the particles and the waves. The full quasilinear resonance condition between particle rigidity and wavenumber reads
\begin{equation}
k = B/(R\mu),
\end{equation}
where $B$ is the magnetic field magnitude and $\mu$ is the pitch-angle cosine, but CSA omits the dependence on $\mu$. As the full resonance condition leads to an ability of low-rigidity particles to resonate with waves generated by high-rigidity particles and vice versa \citep{ng03,ng08}, there may be a difference in the scaling law between the two models. \citet{afanasiev15} presented simulations comparing CSA and a new SOLar Particle Acceleration in Coronal Shocks (SOLPACS) code, where the latter employs the full resonance condition but is restricted to local (one-dimensional) field geometries (i.e., employing a spatially constant mean magnetic field) and parallel shocks in an upstream region of finite size. The results on the scaling of the cutoff energy with shock and plasma parameters between the models are very similar. The cutoff momenta obtained with SOLPACS are about a factor of two lower than those obtained with CSA, but the scaling with the injection efficiency $\epsilon_{\rm inj}=n_{\rm seed}/n$ is practically identical in both cases. Thus, using the results of the global CSA simulation (where the plasma and shock parameters can vary realistically with distance from the Sun) by \citet{vainio14} is justified as far as scaling laws with plasma parameters are concerned.

\section{Results}
\subsection{Fluence spectra of large SEP events}\label{sec:large_sep}
\citet{tylka09} established that the integral rigidity spectra of proton fluences in GLEs can be represented as double power laws with a smooth rollover from one to the other. This Band function \citep{band93} is
\begin{equation}
J(>R)=\begin{cases}
J_0R^{-\gamma_1}\exp(-R/R_0)&R<(\gamma_2-\gamma_1)R_0\equiv R_1\\
J_0R_1^{-\gamma_1}\exp(-R_1/R_0)(R/R_1)^{-\gamma_2}&R\geq R_1\,.
\end{cases}
\end{equation}
The Band function and its derivative are continuous.  \citet{atwell15,atwell16} established the same type of spectral analysis for sub-GLEs, which are less energetic than GLEs, extending into the several hundred MeV range, but without producing detectable levels of secondary atmospheric particles. Neither the CSA nor the SOLPACS simulations reproduce this double power-law spectral form, but the decrease in the fluence beyond the cutoff is typically far more rapid in these simulations. However, as established by \citet{afanasiev14}, an energetically self-consistent model of stochastic acceleration in the turbulent downstream of the shock will harden the spectrum at the highest energies enough to produce a double power law  from an abruptly cutoff spectrum produced at the shock. Thus, we do not consider the Band function to be inconsistent with shock acceleration. We note that another shock acceleration model by \citet{tylka06} also reproduces the double power-law form in a variable magnetic geometry of the shock.

We list the Band fit parameters of the GLEs and sub-GLEs of  SC 23 and 24 in Appendix \ref{sec:app}. The $>$1-GV proton fluences of the sub-GLEs of SCs 23 and 24 are also plotted in Fig.~\ref{fig:glecycles} and given in Table~\ref{tab:fluences} with their error estimates. We compute the fluence values and their uncertainties by generating an ensemble of $10^5$ Band spectra per event, assuming that the parameters are normally distributed with their mean and standard deviation given by the best-fit values and their error, respectively. The $>$1-GV fluence is then computed for each of these Band spectra and the event fluence and its error are computed over the ensemble as the mean and the standard deviation, respectively.

\subsection{Scaling of a large event from SC 23 to SC 24}
Here we  apply the scaling laws to a fluence spectrum of accelerated protons in one of the GLEs of SC 23. This is taken to represent a large event of that cycle, and we  investigate what would happen to the spectrum if the plasma and seed particle densities  were scaled from one cycle to the next as observed on average. We explain the procedure briefly below, but give full details of the analysis in Appendix \ref{sec:app}. 

We take the shock to have a relatively high  but not extreme strength, i.e., we take $r_c=3$ and  thus  $\beta=1.5$. Also, we take the injection rigidity to be constant and consider the same height for the acceleration in both cycles ($r/r_0=$ const.) and this gives the scaling law of the Band function break point rigidity from one cycle to another as
\begin{equation}
R_0 \propto \left(\frac {n_{\rm seed}}{\sqrt{n}}\right)^{1/\beta} = \left(\frac {n_{\rm seed}}{\sqrt{n}}\right)^{2/3}.
\label{eq:R0_scaling}
\end{equation}
The normalization of the spectrum is taken to scale as
\begin{equation}
J_0 \propto n_{\rm seed}.
\end{equation}

\begin{figure}
\centering
\includegraphics [width=0.99\columnwidth, angle=0]{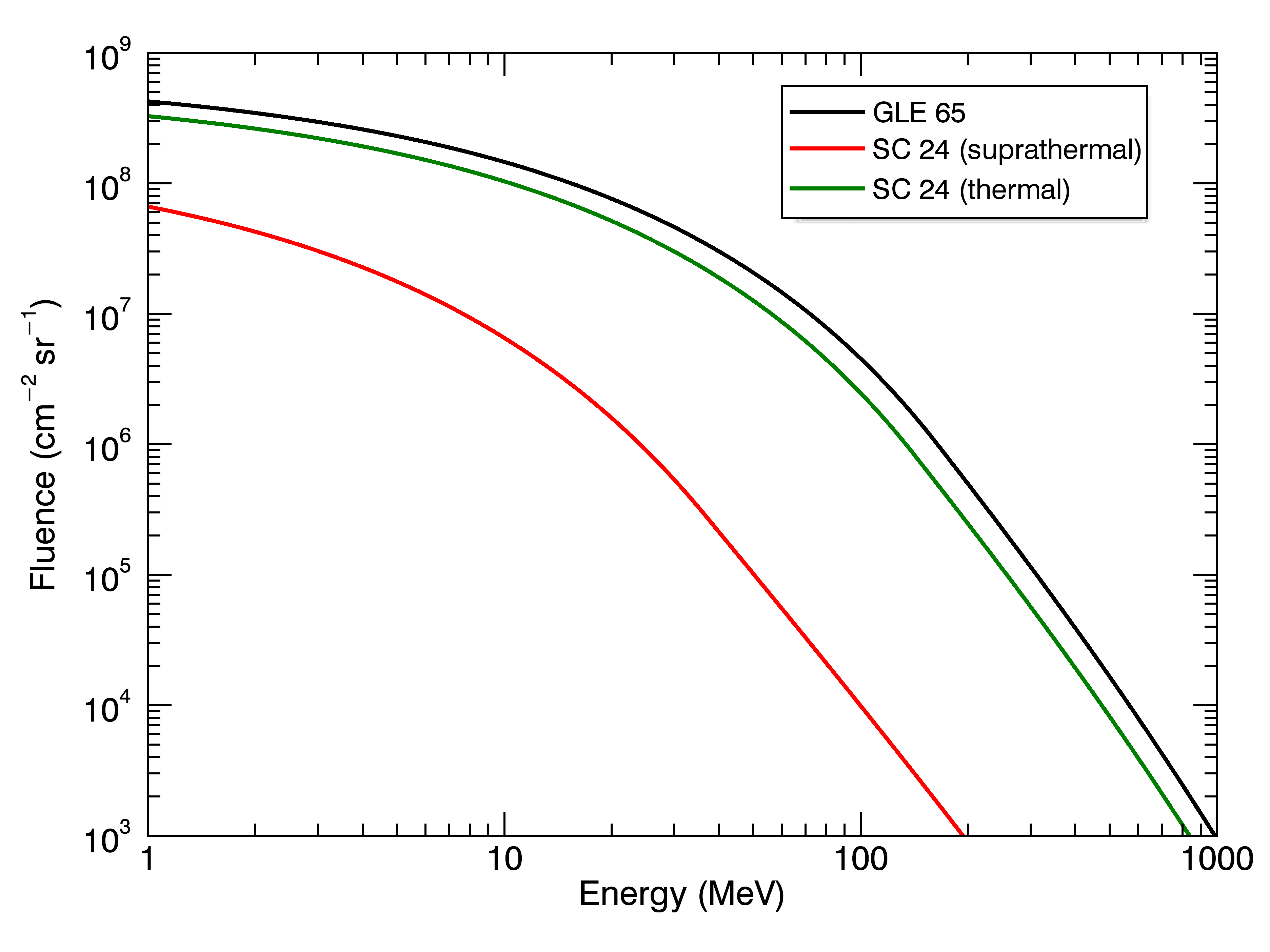}
\caption{ Total integral fluence of the GLE of 28 Oct 2003 (black curve) plotted as a function of energy \citep{tylka09} and the same event with spectral parameters $J_0$ and $R_0$ scaled according to the plasma densities and suprathermal particle densities of the two cycles.}\label{fig:oct-2003_scaled}
\end{figure}

We assume that the GLE 65 on 28 Oct 2003 was accelerated by a CME-driven shock wave and use the simple scaling laws presented above. This yields the spectra presented in Fig.~\ref{fig:oct-2003_scaled}. We take plasma densities to scale as $n^\mathrm{(SC23)}=1.25\,n^\mathrm{(SC24)}$ and  the suprathermal proton densities to scale as $n^\mathrm{(SC23)}_{\rm sth}=3.60\,n^\mathrm{(SC24)}_{\rm sth}$ according to the observations of \citet{gopalswamy14} and \citet{mewaldt15}, respectively, and plot the resulting spectra  assuming $n_{\rm seed}=n$ (green curve) and $n_{\rm seed}=n_{\rm sth}$ (red curve). We note the large difference in the fluence spectra if the seed particles are the suprathermals. This difference arises mainly because of the large difference between the cutoff rigidities of the original and the scaled event, and the decrease in fluence is therefore also more prominent at high than at low rigidities. On the other hand, for thermal particle injection, the differences between the original and scaled events would not be very large.

\subsection{Scaling of cumulative GLE fluence}\label{sec:cumfluence}
\begin{figure}
\centering
\includegraphics [width=0.99\columnwidth, angle=0]{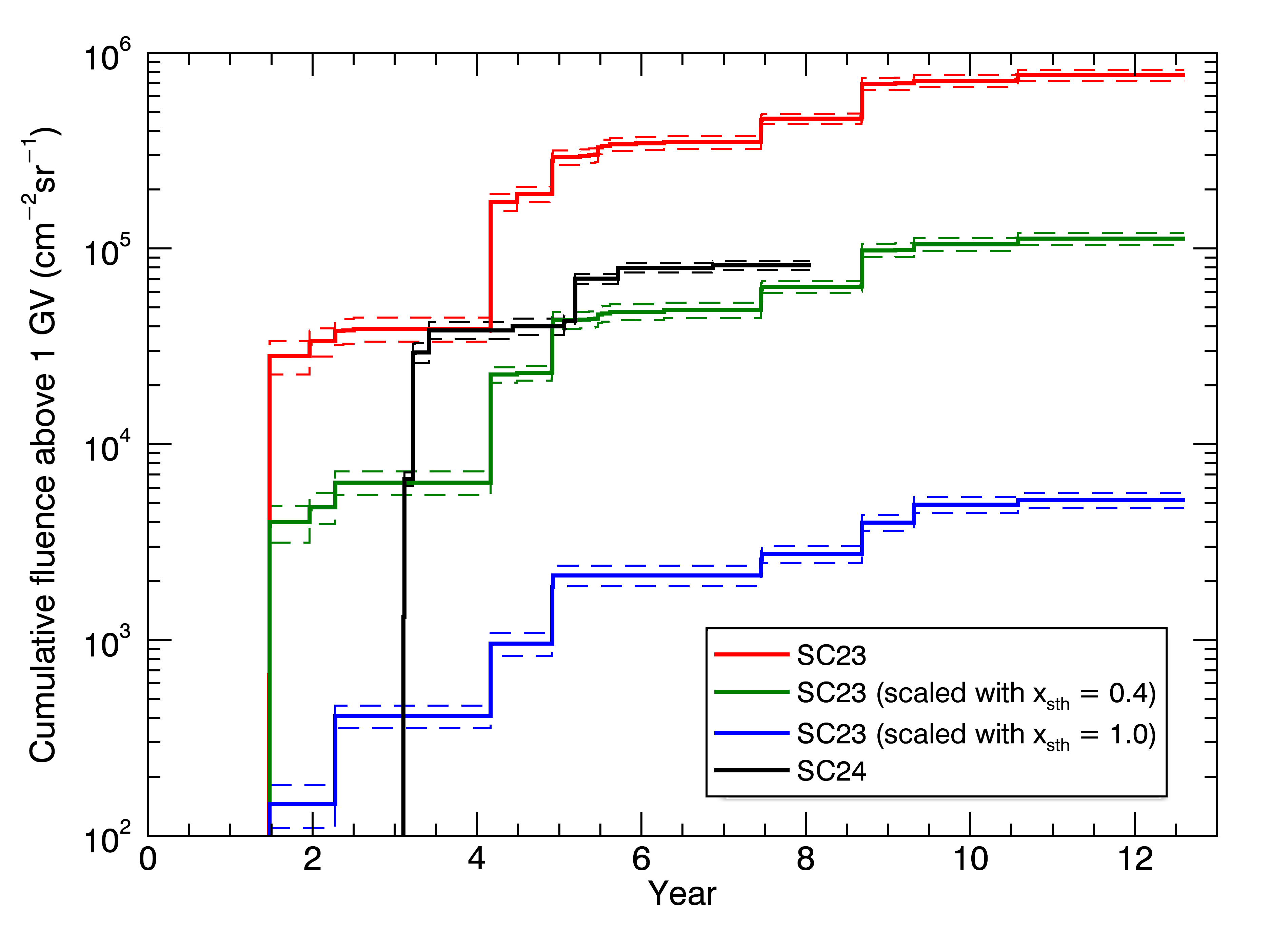}
\caption{ Total cumulative $>$1 GV proton fluence of the GLEs and sub-GLEs of SC 23 (red polyline) plotted as a function of time since the start of the SC; the same GLE fluence plotted by scaling the parameters $J_0$ and $R_0$ to the values representing SC 24 assuming an entirely suprathermal seed population (blue polyline); and the true cumulative GLE fluence of SC 24 (black polyline). The green polyline is the scaled fluence assuming that the seed population consists of 40\% of suprathermals. The 1-$\sigma$ uncertainties are given as dashed polylines around each of the cumulative fluence polylines.}\label{fig:cumulative_glefluence}
\end{figure}

The same scaling law, assuming that the seed population is of suprathermal origin, is then applied to all the GLEs and sub-GLEs of SC 23 to produce a value of the total fluence for the average coronal conditions of SC 24, assuming the same CMEs would occur during this cycle. In Fig.~\ref{fig:cumulative_glefluence} we plot the cumulative proton fluence above 1~GV for SC 23, the cumulative SC 23 GLE fluence but with all events scaled to the coronal conditions of SC 24, and the cumulative GLE and sub-GLE fluence actually observed during SC 24. We note that we include only those events in the scaled fluence that have a $>$1-GV proton fluence $>10^2$~cm$^{-2}\,$sr$^{-1}$, which can be taken as the effective limit of sub-GLEs (see Fig.~\ref{fig:glecycles}). It is evident that the reduction of  seed particles by a factor of 3.6 and a slight decrease in the ambient density by a factor 1.25 are more than enough to explain the lack of high-rigidity particles in the present cycle because the scaled fluence falls well short of the observed value. None of the individual scaled sub-GLE and GLE events are as large as the one observed on 17 May 2012.

We then assume a mixture of thermal and suprathermal protons in the injected population so that a fraction of $x_{\rm sth}$ of the injected ions scales as suprathermal and the rest as the thermal population. Assuming that $x_{\rm sth} = 0.4$, we then compute the scaled parameters of all GLEs and sub-GLEs of SC 23 (see Appendix \ref{sec:app}) and the resulting cumulative $>$1 GV proton fluence from all scaled events that have a fluence  $>10^2$~cm$^{-2}\,$sr$^{-1}$. The resulting cumulative fluence meets the observed SC 24 fluence until the end of the analyzed sub-GLE sample (8  years) within 30\% (Fig.~\ref{fig:cumulative_glefluence}), about 4-$\sigma$.  The results of the fluence scaling analysis for all events are given in Table~\ref{tab:fluences}.

\begin{table*}
\centering
\caption{Calculated  (SC 23 and first 8 years of SC 24) $>$1-GV sub-GLE and GLE fluences along with their scalings (for SC 23 only) to SC 24.}\label{tab:fluences}
\begin{threeparttable}
\begin{tabular}{crrccc} \hline\hline
\vspace{-1.6ex}&&&&\\
& & & Calculated &  \multicolumn{2}{c}{Scaled fluence}\\
Date & Type & Time & fluence &  $x_{\rm sth}=0.4$ & $x_{\rm sth}=1.0$\\
 & &  [a]~~ &[cm$^{-2}\,$sr$^{-1}$]  &[cm$^{-2}\,$sr$^{-1}$]  &[cm$^{-2}\,$sr$^{-1}$] \\\hline
\vspace{-1.6ex}&&&&&\\
\multicolumn{6}{c}{Solar Cycle 23}\\\hline
\vspace{-1.6ex}&&&&&\\
 1997-Nov-04 & s-GLE & $ 1.473$ & $(6.61 \pm 0.71) 10^{+2}                $ & $(6.30 \pm 0.91) 10^{+1}$                 & $(8.20 \pm 3.59) 10^{-1}$                     \\
 1997-Nov-06 &   GLE & $ 1.478$ & $(2.75 \pm 0.54) 10^{+4}                $ & $(3.99 \pm 0.85) 10^{+3}$\tnote{$~\ddagger$} & $(1.45 \pm 0.36) 10^{+2}$\tnote{$~\dagger$}      \\
 1998-May-02 &   GLE & $ 1.963$ & $(4.47 \pm 0.89) 10^{+3}                $ & $(6.50 \pm 1.70) 10^{+2}$\tnote{$~\dagger$}  & $(2.42 \pm 1.30) 10^{+1}$                     \\
 1998-May-06 &   GLE & $ 1.974$ & $(9.21 \pm 0.69) 10^{+2}                $ & $(1.24 \pm 0.14) 10^{+2}$\tnote{$~\dagger$}  & $(3.72 \pm 1.14) 10^{+0}$                     \\
 1998-Aug-24 &   GLE & $ 2.275$ & $(4.27 \pm 0.29) 10^{+3}                $ & $(1.61 \pm 0.14) 10^{+3}$\tnote{$~\dagger$}  & $(2.62 \pm 0.40) 10^{+2}$\tnote{$~\dagger$}      \\
 1998-Sep-30 & s-GLE & $ 2.376$ & $(2.71 \pm 0.26) 10^{+2}                $ & $(3.64 \pm 0.44) 10^{+1}$                 & $(5.27 \pm 0.75) 10^{-1}$                     \\
 1998-Nov-14 & s-GLE & $ 2.500$ & $(8.22 \pm 1.77) 10^{+2}                $ & $(7.96 \pm 4.02) 10^{+1}$                 & $(1.64 \pm 2.20) 10^{+0}$                     \\
 2000-Jul-14 &   GLE & $ 4.164$ & $(1.34 \pm 0.16) 10^{+5}                $ & $(1.63 \pm 0.18) 10^{+4}$\tnote{$~\ddagger$} & $(5.48 \pm 1.14) 10^{+2}$\tnote{$~\dagger$}      \\
 2000-Nov-09 & s-GLE & $ 4.487$ & $(1.62 \pm 0.22) 10^{+4}                $ & $(4.99 \pm 0.72) 10^{+2}$\tnote{$~\dagger$}  & $(1.08 \pm 0.19) 10^{+0}$                     \\
 2001-Apr-03 & s-GLE & $ 4.884$ & $(3.90 \pm 2.22) 10^{+2}                $ & $(3.34 \pm 2.76) 10^{+1}$                 & $(5.11 \pm 7.40) 10^{-1}$                     \\
 2001-Apr-12 & s-GLE & $ 4.909$ & $(1.41 \pm 0.15) 10^{+3}                $ & $(3.20 \pm 0.43) 10^{+2}$\tnote{$~\dagger$}  & $(1.92 \pm 0.27) 10^{+1}$                     \\
 2001-Apr-15 &   GLE & $ 4.917$ & $(8.99 \pm 1.81) 10^{+4}                $ & $(1.68 \pm 0.37) 10^{+4}$\tnote{$~\ddagger$} & $(9.08 \pm 2.13) 10^{+2}$\tnote{$~\dagger$}      \\
 2001-Apr-18 &   GLE & $ 4.925$ & $(1.14 \pm 0.21) 10^{+4}                $ & $(2.82 \pm 0.56) 10^{+3}$\tnote{$~\ddagger$} & $(2.72 \pm 0.67) 10^{+2}$\tnote{$~\dagger$}      \\
 2001-Aug-16 & s-GLE & $ 5.254$ & $(3.27 \pm 0.30) 10^{+3}                $ & $(1.67 \pm 0.21) 10^{+2}$\tnote{$~\dagger$}  & $(9.10 \pm 2.14) 10^{-1}$                     \\
 2001-Sep-24 & s-GLE & $ 5.361$ & $(4.24 \pm 0.38) 10^{+3}                $ & $(1.57 \pm 0.17) 10^{+2}$\tnote{$~\dagger$}  & $(4.05 \pm 0.91) 10^{-1}$                     \\
 2001-Oct-22 & s-GLE & $ 5.437$ & $(1.63 \pm 0.20) 10^{+3}                $ & $(4.84 \pm 0.62) 10^{+2}$\tnote{$~\dagger$}  & $(6.51 \pm 0.99) 10^{+1}$                     \\
 2001-Nov-04 &   GLE & $ 5.473$ & $(2.71 \pm 0.29) 10^{+4}                $ & $(2.00 \pm 0.35) 10^{+3}$\tnote{$~\ddagger$} & $(2.63 \pm 1.01) 10^{+1}$                     \\
 2001-Nov-22 & s-GLE & $ 5.522$ & $(5.76 \pm 5.95) 10^{+3}                $ & $(6.88 \pm 9.12) 10^{+2}$\tnote{$~\dagger$}  & $(1.83 \pm 3.67) 10^{+1}$                     \\
 2001-Dec-26 &   GLE & $ 5.615$ & $(6.81 \pm 1.34) 10^{+3}                $ & $(7.88 \pm 1.76) 10^{+2}$\tnote{$~\dagger$}  & $(1.80 \pm 0.59) 10^{+1}$                     \\
 2002-Apr-21 & s-GLE & $ 5.933$ & $(3.59 \pm 0.34) 10^{+3}                $ & $(1.61 \pm 0.16) 10^{+2}$\tnote{$~\dagger$}  & $(6.88 \pm 0.88) 10^{-1}$                     \\
 2002-Aug-24 &   GLE & $ 6.275$ & $(4.86 \pm 0.99) 10^{+3}                $ & $(8.76 \pm 2.03) 10^{+2}$\tnote{$~\dagger$}  & $(4.78 \pm 1.70) 10^{+1}$                     \\
 2003-Oct-28 &   GLE & $ 7.452$ & $(5.54 \pm 0.74) 10^{+4}                $ & $(5.82 \pm 0.91) 10^{+3}$\tnote{$~\ddagger$} & $(1.18 \pm 0.24) 10^{+2}$\tnote{$~\dagger$}      \\
 2003-Oct-29 &   GLE & $ 7.455$ & $(4.79 \pm 0.37) 10^{+4}                $ & $(7.54 \pm 0.89) 10^{+3}$\tnote{$~\ddagger$} & $(3.22 \pm 0.83) 10^{+2}$\tnote{$~\dagger$}      \\
 2003-Nov-02 &   GLE & $ 7.466$ & $(8.13 \pm 1.48) 10^{+3}                $ & $(1.95 \pm 0.45) 10^{+3}$\tnote{$~\dagger$}  & $(1.67 \pm 0.78) 10^{+2}$\tnote{$~\dagger$}      \\
 2004-Nov-01 & s-GLE & $ 8.465$ & $(1.09 \pm 0.32) 10^{+2}                $ & $(1.30 \pm 0.49) 10^{+1}$                 & $(3.35 \pm 1.83) 10^{-1}$                     \\
 2004-Nov-10 & s-GLE & $ 8.490$ & $(5.72 \pm 1.58) 10^{+2}                $ & $(8.98 \pm 3.06) 10^{+1}$                 & $(3.81 \pm 1.76) 10^{+0}$                     \\
 2005-Jan-17 &   GLE & $ 8.676$ & $(5.97 \pm 0.61) 10^{+3}                $ & $(7.44 \pm 0.83) 10^{+2}$\tnote{$~\dagger$}  & $(2.05 \pm 0.30) 10^{+1}$                     \\
 2005-Jan-20 &   GLE & $ 8.684$ & $(2.27 \pm 0.41) 10^{+5}                $ & $(3.34 \pm 0.63) 10^{+4}$\tnote{$~\ddagger$} & $(1.23 \pm 0.24) 10^{+3}$\tnote{$~\dagger$}      \\
 2005-Jun-16 & s-GLE & $ 9.087$ & $(2.17 \pm 0.47) 10^{+3}                $ & $(3.95 \pm 1.08) 10^{+2}$\tnote{$~\dagger$}  & $(2.17 \pm 0.85) 10^{+1}$                     \\
 2005-Sep-07 & s-GLE & $ 9.314$ & $(2.23 \pm 0.24) 10^{+4}                $ & $(6.75 \pm 1.06) 10^{+3}$\tnote{$~\ddagger$} & $(9.46 \pm 2.61) 10^{+2}$\tnote{$~\dagger$}      \\
 2006-Dec-06 & s-GLE & $10.560$ & $(9.31 \pm 1.01) 10^{+3}                $ & $(9.33 \pm 1.25) 10^{+2}$\tnote{$~\dagger$}  & $(1.58 \pm 0.44) 10^{+1}$                     \\
 2006-Dec-13 &   GLE & $10.579$ & $(4.02 \pm 0.83) 10^{+4}                $ & $(6.40 \pm 1.37) 10^{+3}$\tnote{$~\ddagger$} & $(2.76 \pm 0.66) 10^{+2}$\tnote{$~\dagger$}      \\
\hline
\vspace{-1.6ex}&&&&&\\
\multicolumn{6}{c}{Solar Cycle 24 (first 8 years)}\\\hline
\vspace{-1.6ex}&&&&&\\
2012-Jan-23&s-GLE &  $3.105$ & $(1.31\pm 0.12)\,10^{+3}$ & --- & ---\\
2012-Jan-27&s-GLE &  $3.116$ & $(5.36\pm 0.50)\,10^{+3}$ & --- & ---\\
2012-Mar-07&s-GLE &  $3.225$ & $(2.26\pm 0.33)\,10^{+4}$ & --- & ---\\
2012-Mar-13&s-GLE &  $3.242$ & $(1.44\pm 0.88)\,10^{+2}$ & --- & ---\\
2012-May-17&GLE    &  $3.422$ & $(8.76\pm 1.75)\,10^{+3}$ & --- & ---\\
2013-May-22&s-GLE &  $4.433$ & $(1.86\pm 0.17)\,10^{+3}$ & --- & ---\\
2014-Jan-06&s-GLE &  $5.060$ & $(1.96\pm 0.19)\,10^{+3}$ & --- & ---\\
2014-Jan-07&s-GLE &  $5.062$ & $(6.12\pm 0.80)\,10^{+2}$ & --- & ---\\
2014-Feb-25&s-GLE &  $5.196$ & $(2.76\pm 0.18)\,10^{+4}$ & --- & ---\\
2014-Sep-01&s-GLE &  $5.711$ & $(9.48\pm 0.67)\,10^{+3}$ & --- & ---\\
2015-Oct-29&s-GLE &  $6.869$ & $(2.21\pm 0.27)\,10^{+3}$ & --- & ---\\
\hline
\end{tabular}
  \begin{tablenotes}
     \item[$\dagger$] Exceeds the ``sub-GLE limit'' of $10^2$~cm$^{-2}\,$sr$^{-1}$
     \item[$\ddagger$] Exceeds the ``potential GLE limit'' of $2\cdot 10^3$~cm$^{-2}\,$sr$^{-1}$
  \end{tablenotes}
\end{threeparttable}
\end{table*}

\subsection{Scaling of heavy-ion abundances}\label{sec:abundances}

\begin{figure}
\centering
\includegraphics [width=0.99\columnwidth, angle=0]{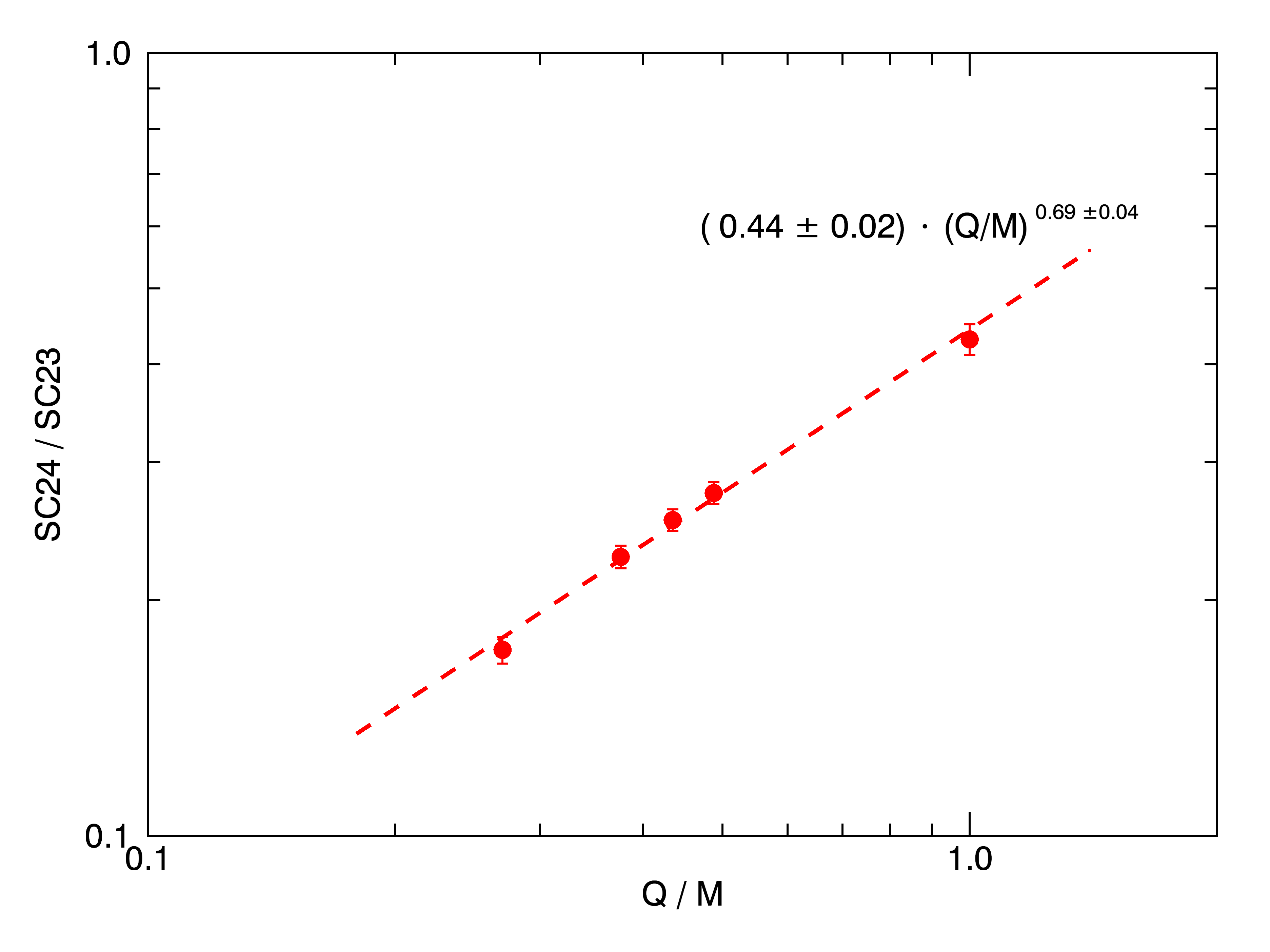}
\caption{Scaling of 10--30 MeV/nuc ion fluences between the two solar cycles (5.8 years from the start of the cycle) with SC 23 GLE spectral parameters scaled to SC 24 using the assumption of 40\% of suprathermals in the injected particle population.}\label{fig:abundance_model}
\end{figure}

The scaling of heavy ion abundances can be done in a similar manner as the modeling of fluence (see Appendix \ref{sec:app} for details). We again take the spectral parameters of the GLEs and sub-GLEs of SC 23 as the starting point and, assuming a mixture of thermal and suprathermal protons in the injected population (see  \S \ref{sec:cumfluence}), we scale the proton spectral parameters to SC 24. For heavy ions, we use the scaling law for the cutoff rigidity of \citet{battarbee11} of $R_c\propto (Q/M)^{-0.2}$. Furthermore, we scale the heavy-ion suprathermal densities relative to H from SC 23 to 24 using the scalings given for O (reduced by a factor of 3.2) and Fe (reduced by a factor of 7.0) by \citet{mewaldt15} and assuming a power-law dependence for $Q/M$ when calculating the other species. The resulting abundance ratio for values of $Q/M$ corresponding to H, C, O, Si, and Fe given by \citet{leske95} is plotted in Fig.~\ref{fig:abundance_model} for the first 5.8 years of the SC to allow a direct comparison with the results of \citet{mewaldt15}. We fit the power law to the points and see that the observational scaling exponent of \citet{mewaldt15} is closely reproduced, when we use the value of $x_{\rm sth}=0.4$ for protons, as in \S \ref{sec:cumfluence}. The predicted overall fluence level in the 10--30 MeV/nuc channel in SC 24 with these assumptions is about a factor of 1.8  higher than the observations \citep{mewaldt15}.

\subsection{Scaling of the number of events}
Finally, we consider the scaling of the number of events detected (Table \ref{tab:fluences}, with $x_{\rm sth}=0.4$). Taking $>$1-GV proton fluence $>10^2$~cm$^{-2}\,$sr$^{-1}$ as the limit of sub-GLE detection, the total number of events in the scaled sample in the first 8 years of the cycle, corresponding to the time period of the last analyzed sub-GLEs of  SC 24, is 20;  the number of detected events in SC 24 for the same time period is 11. If we take $2\cdot 10^3$~cm$^{-2}$~sr$^{-1}$ for the $>$1-GV proton fluence as the limit of GLE detection in  SC 24\footnote{The smallest fluence observed is from GLE 57 (6 May 1998), which is anomalously small for a GLE \citep{thakur16}.}  (see Fig.~\ref{fig:glecycles})  and count the number of scaled GLE fluences that remain above this threshold, we get seven potential GLEs in the scaled sample for the first 8 years of the cycle.

\section{Discussion and conclusions}

We have considered the properties of SEP events over the two most recent SCs. The modeling we  performed  focuses on the role of plasma and seed-particle density in the ambient medium, which according to the simulations \citep{vainio14,afanasiev15} and theory \citep{bell78, vainio07} should be the controlling parameters of the SEP productivity of CME-driven shocks if turbulence around the shock is self-generated in nature. Other parameters that could potentially have effects on the efficiency of shock acceleration can only have  indirect effects on the acceleration process, for example, through the difference in the Mach number of the shock. The Alfv\'en speed may be slightly lower in SC 24 than it was in SC 23 \citep{gopalswamy14}, so if the CME speeds remained the same, the average Mach number of the shock would be higher and this could lead to a larger compression ratio of the shock. Also, close to the critical Mach number ($M_{\rm A}\approx 3$) the injection efficiency of the shock will be strongly increasing, which might have an effect on the moderate shocks. However, as particle events that lead to high SEP fluxes are likely to be due to strong shocks (i.e., their Mach numbers $M_{\rm A}$ are already relatively large), the dependences of $r_c$ and $\epsilon_{\rm inj}$ on $M_{\rm A}$ are expected to be shallow and a small change in Mach number does not make a large difference in the acceleration process. Thus, we expect that on average changes in the Mach number  by approximately 10\% will not affect the results in the largest SEP events that determine the total fluence of the cycle.

We used a  mixture of 40\%  suprathermal and 60\%  thermal protons as the (SC 23) injected particle population to predict the 10--30 MeV/nuc ion abundance reduction from SC 23 to SC 24  as $\propto (Q/M)^\alpha$ with a correct scaling exponent (i.e., $\alpha=0.69\pm 0.04$ when the observed value is 0.66). This choice, however, predicts an overall ion fluence in SC 24 that is higher than observed by a factor of 1.8 for this energy channel. It also predicts that seven events above the potential GLE limit during the first eight years of SC 24 should have been observed. Given the fuzziness in GLE detection (as evidenced by Fig.~\ref{fig:glecycles}), the difference between the prediction and the observational situation during  SC 24 (one official GLE) is not very significant since the total number of events (including sub-GLEs) observed above the potential GLE fluence limit is five. Also the predicted number of events above the sub-GLE limit (20) is similar to the observed number (11) for the first eight years of the cycle, and the predicted cumulative fluence from GLEs and sub-GLEs matches the observed value for the same time range within about 4-$\sigma$. Finally, if  the number of scaled events with a fluence above the largest detected sub-GLE ($2\cdot10^4$ cm$^{-2}\,$sr$^{-1}$) is considered,  we have only one in the whole sample (GLE 70)  occurring 8.68 years after the start of the cycle. The present solar activity level measured with the sunspot number is similar to the conditions of GLE 70, so SC 24 in terms of very energetic SEP events might not be over yet.

As the results of the $Q/M$ scaling of fluences between the cycles were successful, we tested the robustness of the results further. When considering the first 4.8 years or 6.8 years of the cycle, the modeling gives the same scaling within 1-$\sigma$ errors of the parameters. We also performed the analysis of the first 5.8-year scaling by omitting the two largest events from the dataset, which produced the same $Q/M$ scaling exponent but with somewhat lower overall level of SC 24 fluence, i.e., it moves the result closer to the observations. We also made a model run where the actual charge states of the two largest events in the time period were replaced by ones relevant to impulsive flares \citep{luhn87}. The results were practically the same as those presented above. Thus, we consider the presented modeling result to be a robust prediction.

One aspect not considered in the present model is transport. If the transport parameters in the two cycles are different, on average,   the fluence scaling as a function of $Q/M$ will certainly be affected. The resulting pattern of $Q/M$ is difficult to deduce, however,  without detailed event-based modeling. For example, the temporal variation of the Fe/C ratio during large SEP events shows  dependence on the source longitude and inferred temperature \citep{reames16a, reames16b}. Without detailed modeling the net effect in the considered sample of events on the abundance variations is not easy to evaluate. Thus, transport effects constitute an unaccounted error source in our analysis, which may modify the prediction of the $Q/M$ scaling between the cycles. Of course, the variability of the actual coronal and CME parameters with respect to the averages used in the modeling will add another component of uncertainty. The observed daily suprathermal densities, for example,  vary by orders of magnitude \citep{mewaldt15}, which means that basing the analysis on the mere long-term average values is not expected to yield very exact predictions. Other parameters, including those related to the particle transport conditions, have large variations  from event to event   as well. Thus, some of the good agreement achieved in this study is likely to be fortuitous, and one should attempt to study the predicting power of the prevailing densities of the plasma and suprathermals for a large sample of observed large SEP events to obtain further evidence. However, getting to coronal densities from those measured in the solar wind from event to event  might prove to be challenging and is certainly beyond the scope of this paper.

Any results we present here cannot be regarded as conclusive evidence for shock acceleration or---even if shock acceleration is regarded as a proven way of SEP event genesis---as the only possible route to the explanation of the high-energy ion deficit of the present cycle. For example, we have not considered the effect of perpendicular diffusion to the cutoff rigidities, and this should be important at least in quasi-perpendicular shocks \citep[e.g.,][]{zank07}. In that case,  plasma parameters other than density, e.g., the magnetic field intensity, would become more important than in the present modeling; for nearly perpendicular shocks upstream turbulence would be less affected by the accelerated protons since their field-aligned anisotropies driving the streaming instability would be lower and the acceleration rate would scale with the cyclotron frequency. In addition, for quasi-perpendicular shocks, the comparative dearth of suprathermals would likely cause an even greater reduction on high-energy SEP production because of injection threshold requirements \citep{tylka05,tylka06}. Despite its limitations, however, the present study led to intriguing results, lending support to ion acceleration by coronal shocks with proton-generated Alfv\'en waves as the main agent determining the SEP event properties and the seed particle properties as an important factor controlling the highest energies obtained from the process.

\begin{acknowledgements}
This research has received funding from the European Union’s Horizon 2020 research and innovation program under grant agreement No. 637324 (HESPERIA) and from the Academy of Finland, project 267186. O.R. thanks the Vilho, Yrj\"o and Kalle V\"ais\"al\"a foundation for financial support.
\end{acknowledgements}

\bibliographystyle{aa}
\makeatletter
\let\ps@plain=\ps@fancy
\makeatother
\addcontentsline{toc}{chapter}{Bibliography}
\bibliography{ms30547}

\appendix

\section{GLE spectral parameters and their scaling from SC 23 to SC 24}\label{sec:app}
The spectral parameters, i.e., the fits of the observed fluences to the Band function as a function of rigidity for all GLEs with NM increase at stations with high enough cutoff rigidity for spectral determination, were first determined by \citet{tylka09}. We have considered here a slightly revised version of the analysis, which also contains parameters of the latest GLE 71 (2012 May 17). The parameters for all the GLEs of SCs 23 and 24 are given in Table \ref{tab:gle_spectra}.

\begin{table*}
\caption{Spectral parameters for GLE fluences of  SCs 23 and 24. Events labeled with numbers are the official GLEs and the ones below appended with ``{\sc esp}'' are the energetic storm particle components related to these GLEs. The uncertainties are estimated by varying the parameter of interest while holding the others at their best-fit value.}\label{tab:gle_spectra}
\centering
\begin{tabular}{lccccc}\hline\hline
\vspace{-1.6ex}&&&&&\\
GLE   & Date & $J_0\, [\mbox{cm}^{-2}]$ & $\gamma_1$ & $\gamma_2$ & $R_0\, [\mbox{GV}]$ \\\hline
\vspace{-1.6ex}&&&&&\\
55                      &1997-Nov-06&    $(8.15\pm 0.85)\,10^8$& $0.28\pm 0.09$&  $5.38\pm 0.09$& $0.116\pm 0.004$\\
56                      &1998-May-02&    $(8.98\pm 0.76)\,10^6$& $1.31\pm 0.07$&  $6.51\pm 0.55$& $0.196\pm 0.008$\\
57                      &1998-May-06&    $(1.64\pm 0.04)\,10^6$& $1.92\pm 0.02$&  $7.46\pm 0.41$& $0.202\pm 0.003$\\
58                      &1998-Aug-24&    $(2.10\pm 0.05)\,10^5$& $2.98\pm 0.02$&  $5.27\pm 0.34$& $0.677\pm 0.034$\\
58 {\sc esp}    &1998-Aug-24&    $(1.63\pm 0.04)\,10^4$& $5.26\pm 0.03$& $7.74\pm 0.34$&  $0.983\pm 0.060$\\
59                      &2000-Jul-14&    $(2.94\pm 0.30)\,10^9$& $0.51\pm 0.10$&  $7.46\pm 0.14$& $0.123\pm 0.003$\\
59 {\sc esp}    &2000-Jul-14&    $(6.01\pm 0.18)\,10^7$& $3.24\pm 0.05$& $7.85\pm 0.29$&  $0.226\pm 0.005$\\
60                      &2001-Apr-15&    $(5.22\pm 0.54)\,10^7$& $1.39\pm 0.09$&  $5.69\pm 0.08$& $0.260\pm 0.012$\\
61                      &2001-Apr-18&    $(8.39\pm 0.75)\,10^6$& $1.85\pm 0.07$&  $5.02\pm 0.14$& $0.237\pm 0.012$\\
62                      &2001-Nov-04&    $(2.14\pm 0.19)\,10^9$& $0.24\pm 0.07$&  $6.67\pm 0.31$& $0.093\pm 0.003$\\
62 {\sc esp}    &2001-Nov-04&    $(4.78\pm 0.11)\,10^8$& $2.36\pm 0.04$& $11.2\pm 0.3$&           $0.129\pm 0.002$\\
63                      &2001-Dec-26&    $(2.17\pm 0.19)\,10^7$& $1.81\pm 0.07$&  $7.86\pm 0.34$& $0.180\pm 0.006$\\
64                      &2002-Aug-24&    $(5.06\pm 0.46)\,10^6$& $2.36\pm 0.08$&  $6.70\pm 0.33$& $0.225\pm 0.010$\\
65                      &2003-Oct-28&    $(8.44\pm 0.78)\,10^9$& $0.01\pm 0.08$&  $6.48\pm 0.25$& $0.089\pm 0.003$\\
65 {\sc esp}    &2003-Oct-28&    $(1.12\pm 0.04)\,10^8$& $2.81\pm 0.05$& $8.92\pm 0.10$&  $0.171\pm 0.003$\\
66                      &2003-Oct-29&    $(7.62\pm 0.18)\,10^7$& $2.04\pm 0.02$&  $6.86\pm 0.30$& $0.206\pm 0.004$\\
67                      &2003-Nov-02&    $(2.27\pm 0.20)\,10^6$& $3.50\pm 0.07$&  $7.01\pm 0.52$& $0.321\pm 0.017$\\
68                      &2005-Jan-17&    $(3.51\pm 0.11)\,10^7$& $2.65\pm 0.05$&  $8.29\pm 0.11$& $0.162\pm 0.003$\\
69                      &2005-Jan-20&    $(3.81\pm 0.29)\,10^8$& $0.72\pm 0.07$&  $5.78\pm 0.06$& $0.204\pm 0.007$\\
70                      &2006-Dec-13&    $(1.33\pm 0.14)\,10^8$& $1.05\pm 0.09$&  $5.80\pm 0.09$& $0.177\pm 0.007$\\
71                      &2012-May-17&    $(1.03\pm 0.11)\,10^7$& $1.36\pm 0.09$&  $6.96\pm 0.97$& $0.219\pm 0.008$\\\hline
\end{tabular}
\end{table*}

\begin{figure}[h]
\centering
\includegraphics [width=0.99\columnwidth, angle=0]{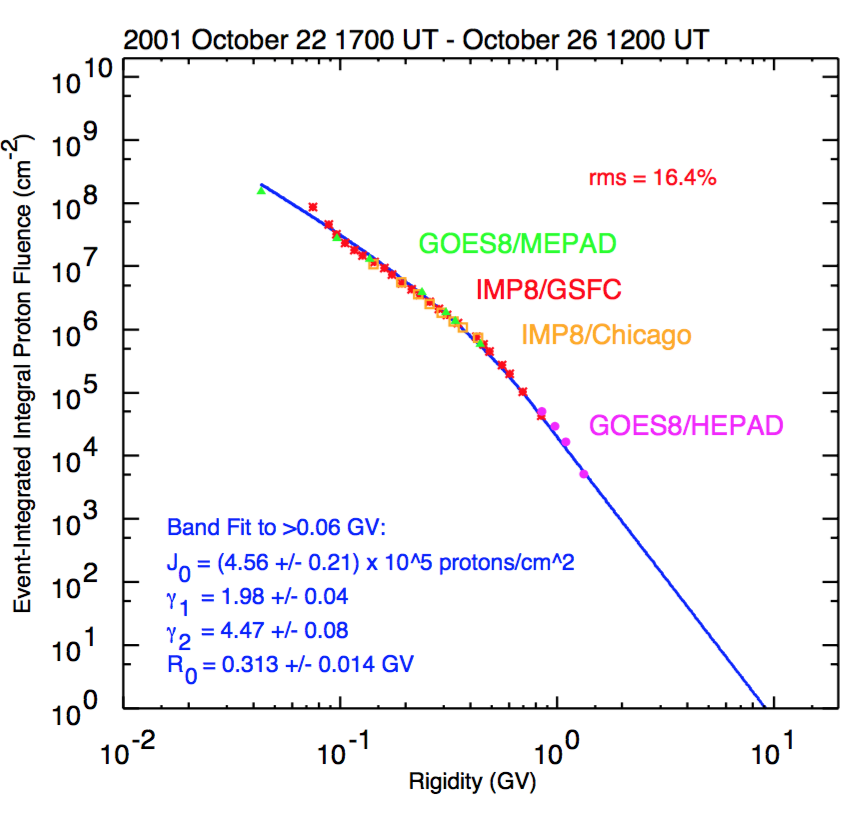}
\caption{ Sub-GLE of 22 Oct 2001  fitted with IMP-8 data (GSFC and Chigaco instruments) and compared with GOES fluence data used throughout the rest of the analysis.}\label{fig:fluencecomp}
\end{figure}

In addition to GLEs, we also consider the SEP events extending to several hundred MeVs but not producing a GLE. We made this choice because the conditions under which a particle event becomes a GLE do not depend on a simple fluence or flux limit observed during the event. The inclusion of sub-GLE events in this case means that we also consider  the events that have been observed by the GOES/HEPAD detector. In one case, the sub-GLE of 22 October 2001, the identified event was analyzed using IMP-8 data instead of GOES for the purposes of cross-checking. The overall fluence spectrum fitted to the data of IMP-8 instruments closely agrees with the GOES data points  (Fig.~\ref{fig:fluencecomp}). The list of Band fit parameters for the sub-GLEs of SC 23 and 24 are given in Table \ref{tab:sub-gle_spectra}.

\begin{table*}
\caption{Spectral parameters for sub-GLE fluences of  SCs 23 and 24. The uncertainties are estimated by varying the parameter of interest while holding the others at their best-fit value.}\label{tab:sub-gle_spectra}
\centering
\begin{tabular}{ccccc}\hline\hline
\vspace{-1.6ex}&&&&\\
Date & $J_0\, [\mbox{cm}^{-2}]$ & $\gamma_1$ & $\gamma_2$ & $R_0\, [\mbox{GV}]$ \\\hline
\vspace{-1.6ex}&&&&\\
1997-Nov-04 & $(2.97\pm 0.09)\,10^{6}$ & $1.62\pm 0.02$ & $8.89\pm 0.64$ & $0.170\pm 0.003$ \\
1998-Sep-30 & $(3.64\pm 0.13)\,10^{5}$ & $3.86\pm 0.04$ & $10.9\pm 0.1$ & $0.214\pm 0.004$ \\
1998-Nov-14 & $(2.22\pm 0.06)\,10^{7}$ & $1.39\pm 0.02$ & $8.20\pm 0.88$ & $0.128\pm 0.002$ \\
2000-Nov-09 & $(3.64\pm 0.11)\,10^{10}$ & $0.19\pm 0.02$ & $10.9\pm 0.2$ & $0.082\pm 0.001$ \\
2001-Apr-03 & $(1.35\pm 0.04)\,10^{9}$ & $0.61\pm 0.03$ & $7.98\pm 0.65$ & $0.066\pm 0.001$ \\
2001-Apr-12 & $(3.63\pm 0.11)\,10^{5}$ & $2.42\pm 0.02$ & $6.52\pm 0.03$ & $0.331\pm 0.011$ \\
2001-Aug-16 & $(5.47\pm 0.16)\,10^{8}$ & $0.23\pm 0.02$ & $9.12\pm 0.25$ & $0.105\pm 0.001$ \\
2001-Sep-24 & $(4.71\pm 0.15)\,10^{8}$ & $1.93\pm 0.03$ & $12.3\pm 0.3$ & $0.110\pm 0.001$ \\
2001-Oct-22 & $(4.56\pm 0.21)\,10^{5}$ & $1.98\pm 0.04$ & $4.47\pm 0.08$ & $0.313\pm 0.014$ \\
2001-Nov-22 & $(2.44\pm 0.10)\,10^{11}$ & $0.04\pm 0.05$ & $6.33\pm 0.54$ & $0.038\pm 0.001$ \\
2002-Apr-21 & $(6.13\pm 0.22)\,10^{8}$ & $1.36\pm 0.03$ & $10.7\pm 0.1$ & $0.105\pm 0.001$ \\
2004-Nov-01 & $(1.04\pm 0.04)\,10^{8}$ & $0.28\pm 0.02$ & $6.17\pm 0.29$ & $0.068\pm 0.001$ \\
2004-Nov-10 & $(2.41\pm 0.10)\,10^{8}$ & $1.13\pm 0.02$ & $5.99\pm 0.22$ & $0.065\pm 0.001$ \\
2005-Jun-16 & $(1.19\pm 0.05)\,10^{8}$ & $0.04\pm 0.03$ & $4.36\pm 0.21$ & $0.090\pm 0.001$ \\
2005-Sep-07 & $(5.94\pm 0.18)\,10^{6}$ & $3.10\pm 0.02$ & $5.55\pm 0.24$ & $0.317\pm 0.010$ \\
2006-Dec-06 & $(4.18\pm 0.13)\,10^{7}$ & $2.32\pm 0.02$ & $8.97\pm 0.36$ & $0.170\pm 0.003$ \\
2012-Jan-23 & $(1.96\pm 0.11)\,10^{5}$ & $4.96\pm 0.22$ & $9.65\pm 0.38$ & $0.403\pm 0.011$ \\
2012-Jan-27 & $(2.22\pm 0.07)\,10^{7}$ & $1.94\pm 0.03$ & $6.04\pm 0.10$ & $0.161\pm 0.003$ \\
2012-Mar-07 & $(1.53\pm 0.04)\,10^{9}$ & $1.06\pm 0.02$ & $7.73\pm 0.42$ & $0.112\pm 0.001$ \\
2012-Mar-13 & $(3.33\pm 0.10)\,10^{7}$ & $1.30\pm 0.02$ & $7.45\pm 0.75$ & $0.087\pm 0.001$ \\
2013-May-22 & $(5.07\pm 0.14)\,10^{5}$ & $3.77\pm 0.04$ & $8.25\pm 0.41$ & $0.325\pm 0.009$ \\
2014-Jan-06 & $(6.35\pm 0.19)\,10^{6}$ & $1.08\pm 0.01$ & $8.34\pm 0.44$ & $0.180\pm 0.003$ \\
2014-Jan-07 & $(5.05\pm 0.17)\,10^{6}$ & $3.06\pm 0.05$ & $11.7\pm 1.0$ & $0.154\pm 0.003$ \\
2014-Feb-25 & $(1.74\pm 0.06)\,10^{6}$ & $3.46\pm 0.03$ & $5.49\pm 0.09$ & $0.620\pm 0.022$ \\
2014-Sep-01 & $(1.87\pm 0.07)\,10^{6}$ & $1.54\pm 0.04$ & $4.57\pm 0.27$ & $0.363\pm 0.008$ \\
2015-Oct-29 & $(1.10\pm 0.05)\,10^{9}$ & $0.51\pm 0.02$ & $4.51\pm 0.04$ & $0.048\pm 0.001$\\\hline
\end{tabular}
\end{table*}

To scale the events from SC 23 to SC 24, we adopt the following scheme. Particles injected into the shock acceleration process are assumed to be a mixture of thermal and suprathermal ions. For protons, the fraction of suprathermals in the injected population in SC 23 $x_{\rm sth}$ is assumed to be constant for all events. Thermal proton density from one SC to the next is assumed to scale as 
\begin{equation}
n^\mathrm{(SC23)}=1.25\, n^\mathrm{(SC24)}
\end{equation}
and the suprathermal proton density as\begin{equation}
n_{\rm sth}^\mathrm{(SC23)}=3.6\, n_{\rm sth}^\mathrm{(SC24)}
,\end{equation}as found by \citet{mewaldt15}. Thus, the injected proton density is assumed to scale as
\begin{equation}
n_{\rm seed,p}^\mathrm{(SC24)} = [(1-x_{\rm sth})/1.25+x_{\rm sth}/3.6]\,n_{\rm seed,p}^\mathrm{(SC23)}.
\end{equation}
This is used to fix the scaling of the cutoff rigidity of protons from SC 23 to SC 24 using Eq.~(\ref{eq:R0_scaling})
\begin{eqnarray}
R_{0,\rm p}^\mathrm{(SC24)} & = &\left(\frac{n_{\rm seed,p}^\mathrm{(SC24)}}{n_{\rm seed,p}^\mathrm{(SC23)}}\right)^{2/3}
\left(\frac{n^\mathrm{(SC23)}}{n^\mathrm{(SC24)}}\right)^{1/3}R_{0,\rm p}^\mathrm{(SC23)} \nonumber\\
 & = & [ (1-x_{\rm sth})/1.25+x_{\rm sth}/3.6]^{2/3}\,1.25^{1/3}R_{0,\rm p}^\mathrm{(SC23)}.
\end{eqnarray}

We regard the GLE and sub-GLE spectra to be primary proton spectra outside the Earth's magnetosphere. To scale the spectral parameters to other species, we assume that (i) the abundance ratios of elements do not vary from one event to another; (ii) the cutoff rigidities ($R_0$) scale as \citep{battarbee11} \begin{equation}
R_0=R_{0,\rm p}(Q/M)^{-0.2};
\end{equation}and (iii) the charge states of the analyzed elements (H, C, O, Si, Fe) correspond to those given by \citet{leske95} and do not vary from event to event.

For thermal minor ions, we assume that the density scaling law from one cycle to the next is the same for all species. Heavier suprathermal ions are assumed to be more suppressed than protons. We assume that the scaling follows $n_{\rm sth}^\mathrm{(SC23)}/n_{\rm sth}^\mathrm{(SC24)}\propto (Q/M)^{-\alpha}$ and we fix the spectral exponent using the observed values for O and Fe \citep{mewaldt15}, giving $\alpha=1.65$. Thus, the applied species-dependent scaling of spectral normalization is
\begin{equation}
J_0^\mathrm{(SC24)} =  [(1-x_{\rm sth})/1.25+(Q/M)^{1.65}\,x_{\rm sth}/3.6]\,J_0^\mathrm{(SC23)}.
\end{equation}

We then calculate the fluences of H, C, O, Si, and Fe for all GLEs (Table \ref{tab:gle_spectra}) and the sub-GLEs (Table \ref{tab:sub-gle_spectra}) of  SC 23,  and their counterparts scaled to SC 24. This is done with a Monte Carlo analysis that takes into account the error limits of the spectral parameters. Assuming that the Band-fit spectral parameters are normally distributed we generate an ensemble of $10^5$ spectra per each observed event, and determine the fluence for all of them. The event fluence and its error are then obtained as the mean and standard deviation over this ensemble. Summing  the events in each cycle and taking the ratios of cumulative fluences for both cycles then gives us the scaling of ion fluxes from one cycle to the next for each assumed value of $x_{\rm sth}$.

\end{document}